\documentclass[article]{revtex4}
\usepackage[dvips]{graphicx,color}
\usepackage{epsfig}
\usepackage{epstopdf}
\usepackage{ifpdf}
\usepackage{latexsym}
\usepackage{slashed}
\usepackage{subfigure}
\usepackage{amsmath}
\usepackage{amssymb}
\usepackage{amsfonts}
\bibliography{references}
\begin{document}
\title {Numerical Studies of    
Particle Laden  Flow in Dispersed Phase} 

\author{R. Dutta\footnote{Electronic Address:rdutt@orca.st.usm.edu}}
\address{Physics department, 
The Ohio State University, Columbus, Ohio
}
\author{Shar Sajjadi\footnote{Electronic Address:Shared.Sajjadi@erau.edu}}
\address{Dept. of Mathematics, Embry Riddle University, 
Daytona Beach, Florida}

\maketitle
\medskip
{\it Proceedings of Summer Program 2010} 
\section{abstract}
 To better understand the hydrodynamic flow behavior in turbulence, Particle-Fluid flow 
 have been studied using our Direct Numerical(DNS) based software DSM on MUSCL-QUICK 
and finite volume algorithm. The particle flow was studied using Eulerian-Eulerian 
Quasi Brownian Motion(QBM) based approach. 
 The dynamics is shown for various particle sizes 
which are very relevant to spray mechanism for Industrial applications and Bio medical applications.  
\section{Introduction and Motivation}
Particle laden flows are of great interest in Chemical,
Spray Industrial applications or Biomedical applications. 
Knowledge of particle transport and concentration properties are crucial 
for experimental design of 
such applications. Numerical simulation coupling Lagrangian tracking  in 
 discrete carrier phase of
turbulence with Direct Numerical Simulation(DNS) simulation phase 
 provides a robust tool to investigate such flows. 
Many applications 
of this type of flow are aerosol[Nitrogen and other post combustion particles]
particle flow studies in post combustion chamber for 
the design and performance 
optimization of aviation engine and also for  spray mechanism in nano particle spray Industries. In studies 
of arterial blood flow, these simulations help to visualize the blood flow in cardio-vascular region. 
Numerical Simulation coupling Lagrangian approach of particle tracking along with 
DNS(Direct Numerical Simulation) phase of simulation provides a powerful tool for such flow such 
in any geometry. \\
 In all Industrial applications of the boundary wall problem, 
boundaries are  not smooth in true sense and the presence of roughness cause additional energy dissipation 
enhancing mixing in the particle-fluid mixture. Most commercial software like Fluent and CFX
 are unable to incorporate such physics. There are two commonly used methods to simulate fluid-particle
flows:Eulerian-Lagrangian and Eulerian-Eulerian, and both of these have been used to model the settling 
of particles in an in compressible fluids. The Eulerian-Lagrangian technique treats the fluid 
as a continuous medium described by Navier-Stokes equation modified to account for the fluid-particle 
interaction where particles are considered as point in fluid such that Newton's second law can be applied
separately to each particle to track its motion in a Lagrangian frame of reference. Particle-particle
interaction and particle-fluid interactions are modeled for each particle. When particle numbers become large, 
particle particle interaction and turbulence modifications become expensive to study such kind of flows.
This even sometimes becomes computationally challenging to retrieve data \cite{kosinski}. 
These effects become important in the calculation which makes DNS simulation approach to become numerically 
Numerical computation based on separate Eulerian balance can provide very good alternative approach
to such problems which are numerically not so expensive in comparison to grid size. 
Eulerian approach is based on separate balance equation for both phases through inter phase 
coupling terms.
Such Eulerian-Eulerian DNS approach have been validated for the case of particles with low 
inertia which follow their carrier fluid almost instantaneously due to their small response time 
compared to integral time scale of turbulence[Druzini et al] \cite{druzini}.
In case of inertial particles, where particle response time scale is comparable to integral time scale,
 additional effects
have to be taken into account.  
As pointed out by Fevrier et al \cite{fevrier1}  \cite{fevrier2} calculations, 
particle phase transport
equations must take account all dispersion effects due to local random motion 
which is induced by particle-particle interaction and particle-turbulence interaction. In fact, complex 
nonlinear particle-particle interaction exists in the vicinity of wall. \\
Following Fevrier et al \cite{fevrier1}, a conditional average of all dispersed phases with
respect to the carrier phase flow allows the derivation of instantaneous mesoscopic 
particle fields and instantaneous Eulerian balance equation which can be calculated by 
taking into account for the effect of random motion. 
The detail calculation of energy dissipation parameter also takes into account of random motion
of the dispersed phase. \\
Using forced isotropic turbulence simulations, Fevrier et al \cite{fevrier2} showed that 
uncorrelated quasi Brownian motion of the particles increases with inertia(Stokes Number). \\
In cases where particle relaxation time is comparable to Lagrangian integral time scale,
kinetic energy of the quasi Brownian motion is about 30\% of total kinetic energy 
of the dispersed phase. \\
The importance of Quasi Brownian Motion(QBM) is 
illustrated in a preliminary test in case of decaying homogeneous isotropic turbulence. 
The Eulerian model is then applied to the experimental case of Snyder \& Lumley \cite{snyder}
which has been previously simulated using Lagrangian tracking approach by  Elghobashi \& 
Trusdel \cite{elghobashi}. 
This allows us to compare results of simulation with Lagrangian  approach and also with the 
experimental data.  The next section will illstrate the Eulerian model and related calculation. 
\section{The Eulerian Model}
 Eulerian equations for the dispersed phase may be derived using the methods
which consists of volume filtering of the separate, local instantaneous 
equations accounting for the inter facial jump conditions [Druzini et al\cite{druzini}].
Such an averaging approach is very restrictive because particle size and 
inter particle distance have to be smaller than the smallest turbulence scale. \\
A different approach in the framework of kinetic theory
is the statistical approach. In analogy to the derivation of Navier Stokes equation
by non equilibrium statistics by Chapman \& Cowling \cite{chapman}, a point probability
density function(pdf) $ {f_{\rm p}}{\rm {(1)}}(c_{\rm p}; x_{\rm p},t) $ defining  
the local instantaneous probable number of particle centers with the given translation 
velocity $u_{\rm p}=c_{\rm p} $, is defined. This function obeys Boltzmann type of kinetic equation
which accounts for the momentum exchange with carrier fluid and other inter particle
forces and external nonlinear force field. \\
Reynold-averaged Transport equations for the first moment such as particle concentration,
mean velocity and particle kinetic stress can be derived directly from  this probability
density function of Kinetic equation [Simonin et. al.\cite{simonin}]. \\
To derive local instantaneous Eulerian Equation in dilute flows(without turbulence modifications 
of the particles), Fevrier et al \cite{fevrier2} proposed an approach that uses averaging over all 
dispersed phases over single carrier phase condition. Such an averaging leads to conditional 
particle dispersion function(pdf) for the dispersed phase defined by,  
\begin{equation}
\check{{f_{\rm p}}^{\rm {(1)}}} (c_{\rm p};{\bf x}, t,{\it H_{\rm f}})= \langle {W_{\rm p}}^{\rm 1} 
(c_{\rm p};x,t \vert H_{\rm f} \rangle
\end{equation}
The $W_{\rm p} $ 's are the position and velocities of corresponding particle with time.  
 With this definition, one may define local instantaneous particulate 
velocity field defined as "Mesoscopic Eulerian Particle Velocity Field". This field is 
obtained by averaging discrete particle velocities measured at a particular position and time
for all particle-flow realizations and at given carrier-phase realization. \\
Such an averaging leads to a conditional "pdf" for the dispersed phase,
\begin{equation}
\check{u_{\rm p}}({\bf u}, t,{\it H_{\rm f}}) = \cfrac{1}{{\check{n_{\rm p}}}^{\rm {(1)}}}  
\int c_{\rm p} \check{f_{\rm p}}^{\rm {(1)}} ( c_{\rm p}; {\bf x},t, {\it H_{\rm f}}) dc_{\rm p} 
\end{equation}
Here
\begin{equation}
   \check{n_{\rm p}}^{\rm {(1)}}  = \int \check{f_{\rm p}}^{\rm {(1)}}(c_{\rm p};
{\bf x}, t,{\it H_{\rm f}})dc_{\rm p} 
\end{equation}
is the mesoscopic particle number density. Application to the conditional averaging
procedure to the kinetic equation governing the particle pdf directly
leads to the transport equation for the first moments of number density
and mesoscopic Eulerian velocity. \\
\begin{equation}
\cfrac{\partial}{\partial t} \check{n_{\rm p}} + \cfrac{\partial}{\partial x_{\rm i} } 
\check{n_{\rm p}} \check{u_{\rm {p,i}}}=0  
\end{equation}
\vskip 1pt
\begin{equation}
\check{n_{\rm p}} \cfrac{\partial}{\partial t} \check{u_{\rm {p,i}}} + 
\check{n_{\rm p}} \check{u_{\rm {p,j}}}\cfrac{\partial}
{\partial x_{\rm j}} \check{u_{\rm {p,i}}} = - \cfrac{ \check{n_{\rm p}}}{\check{
\tau_{\rm p}}^{\rm F} } [ \check{u_{\rm {p,i}}} - u_{\rm {f,i}} ] 
- \cfrac{\partial}{\partial x_{\rm j}} \check{n_{\rm p}} {\delta} \check{\sigma_{\rm {p,i,j}}}
+ \check{n_{\rm p}}g_{\rm i}  
\end{equation}
Here $ \delta \check{\sigma_{\rm {p,i,j}}}$ is the mesoscopic kinetic stress 
tensor of the particle Quasi Brownian
velocity distribution. Our calculation showed that this term is non negligible 
for inertial particles in turbulent flow. 
\subsection{The Stress Tensor of Quasi Brownian Motion(QBM)}
The stress term in eqn(5) arises from the ensemble average of the non linear term
in the transport equation for particle momentum,
\begin{equation}
\begin{split}
n_{\rm p} {\partial} \check{\sigma_{\rm {p,ij}}} = \int (c_{\rm p} - \check{u_{\rm {p,i}}})(c_{\rm {p,j}}
- \check{u_{\rm {p,j}}}) \check{f_{\rm p}}^{\rm {(1)}} (c_{\rm p};{\bf x},t,{\it H_{\rm f}})dc_{\rm p} \\
= \check{n_{\rm p}}{\delta}u_{\rm {p,i}}\widehat{\delta u_{\rm {p,i}}}
\end{split}
\end{equation}
When Euler or Navier Stokes equation is derived from kinetic gas theory, the trace of 
$ {\delta}u_{\rm p} \widehat{\delta}u_{\rm {p,j}} $ is interpreted as 
temperature(ignoring Boltzman constant and molecular mass)
 and related to 
pressure by equation of state. \\
In Euler or Navier Stokes equation, temperature is defined as the uncorrelated part of
kinetic energy. And uncorrelated part of kinetic energy is defined as, 
\begin{equation}
{\partial}{q_{\rm p}}^{\rm 2} = \cfrac{1}{2} {\delta}u_{\rm {p,i}}\widehat{\delta u_{\rm {p,i}}}
\end{equation}
In analogy to Euler or Navier Stokes equation, product of uncorrelated kinetic energy 
and particle number density is defined as uncorrelated Quasi Brownian Pressure(QBP) as,
\begin{equation}
\check{P_{\rm p}} = \check{n_{\rm p}}\cfrac{2}{3} {\delta}{q_{\rm p}}^{\rm 2}
\end{equation}
 When particle number density becomes non uniform, as in the case 
of compressible gas, the pressure tends to homogenize particle
number density.  \\
The non diagonal part of stress tensor can be identified in analogy to Navier Stokes 
Equation as viscous term($\Theta_{\rm {ij}}$). The momentum transport
equation (5)  becomes, 
\begin{equation}
\check{n_{\rm p}} \cfrac{\partial}{\partial t} \check{u_{\rm {p,i}}} + 
\check{n_{\rm p}} \check{u_{\rm {p,j}}} 
\cfrac{\partial}{\partial x_{\rm j}} \check{u_{\rm {p,i}}}
= - \cfrac{\check{n_{\rm p}}}{\check{\tau_{\rm p}}^{\rm F}} [ \check{u_{\rm {p,i}} }
- \check{u_{\rm {f,i}}} ] - \cfrac{\partial}{\partial x_{\rm i}} \check{P_{\rm p}} 
+ \cfrac{\partial}{\partial x_{\rm j}} {\theta}_{\rm {i,j,p}} + n_{\rm {p}}g_{\rm i}
\end{equation}
Moreover, it can be shown mathematically that equation (9) without pressure like term leads to
nonphysical solution.  
\subsection{ Simulation with and without QBM}
Preliminary simulations were performed without any stress term related to QBM. 
Particles tend to accumulate rapidly in smaller region causing unnecessary high particle
number density. This causes numerical simulation to fail. In order to ensure that failure
is not caused by numerical problem, different simulations with different Reynold Number 
was carried out showing same result. \\
Simulations with quasi Broqnian pressure(QBP) and without quasi Brownian viscous stress
were done on the above test cases. Fevrier et al \cite{fevrier2} measured in forced 
homogeneous isotropic turbulence, a mean quasi Brownian kinetic energy $\partial q^{\rm 2}$
proportional to the mean mesoscopic kinetic energy 
$ {\check{q_{\rm p}}}^{\rm 2} = \cfrac{1}{2} \langle \check{u_{\rm {p,i}}}
{\check{u_{\rm {p,i}}}} \rangle $ 
with a proportionality coefficient depending on the Stokes number. \\
The relation between the quasi Brownian kinetic energy and mean resolved kinetic
energy is used,
\begin{equation}
{\delta} {q_{\rm p}}^{\rm 2} =5{\check{q_{\rm p}}}^{\rm 2} 
\end{equation}
Such a QBP model allows all the test cases that failed without  quasi Brownian stress term to 
simulate. Compared to the value obtained by Fevrier et al \cite{fevrier2}, 
relation (10) overestimates the
quasi Brownian kinetic energy. When viscous stress term is incorporated in the calculation, 
it reduces the pressure. \\
In order to quantify the effect of particle confinement, the normalized variance
of particle number density is introduced,
\begin{equation}
g(r,t) = \cfrac{ \langle n(x,t)n(x+r,t)\rangle }{ {\langle(n(x,t)\rangle}^{\rm 2}}
\end{equation}
\begin{figure}[h]
\includegraphics[scale=0.50]{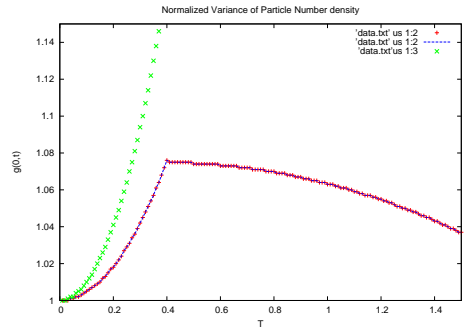}
\caption{Fig 1. Normalized variance distribution of particle number density  without QBP(green color) 
 and with
QBP(red color)} 
\end{figure}
Figure 1 compares time evolution of $g(0,t)$ from simulations with and without QBP. 
The quasi Brownian pressure is found to affect the particle conglomeration 
or segregation significantly. \\
Figure 2 shows kinetic energy spectra of the carrier phase and the dispersed phase
with and without QBP. Our simulations show same behavior of simulations that 
without QBP case, particle kinetic energy of small scale turbulence becomes larger
than the carrier phase in contrast to the results of Fevrier et al {\cite{fevrier2}. 
 This is probably due to nonphysically large particle accumulation in the
 carrier phase of turbulent flow which is the region 
of high shear strain and low vorticity. The vorticity development can be another useful 
tool to study such flows. Figure 5 shows the vorticity development for the 
carrier phase as well as dispersed phase.
 In presence of QBP contribution limiting particle
segregation, particle enstrophy behavior shows similar effect. 
Figure 3 shows particle enstrophy distribution
with and without QBP.  The quasi Brownian viscous 
term plays a major role by inducing a strong small scale strong dissipation 
effect in addition to drag force. 
Figure 3 shows the effect by characterizing the temporal increase of enstrophy function. 
\begin{figure}[ht]
\subfigure[Fig 2. non dimensional energy 
spectra of $\dotplus(red)$ carrier phase $\diamond(pink)$ without QBM and $\times(blue)$ with QBM]{
\includegraphics[scale=0.45]{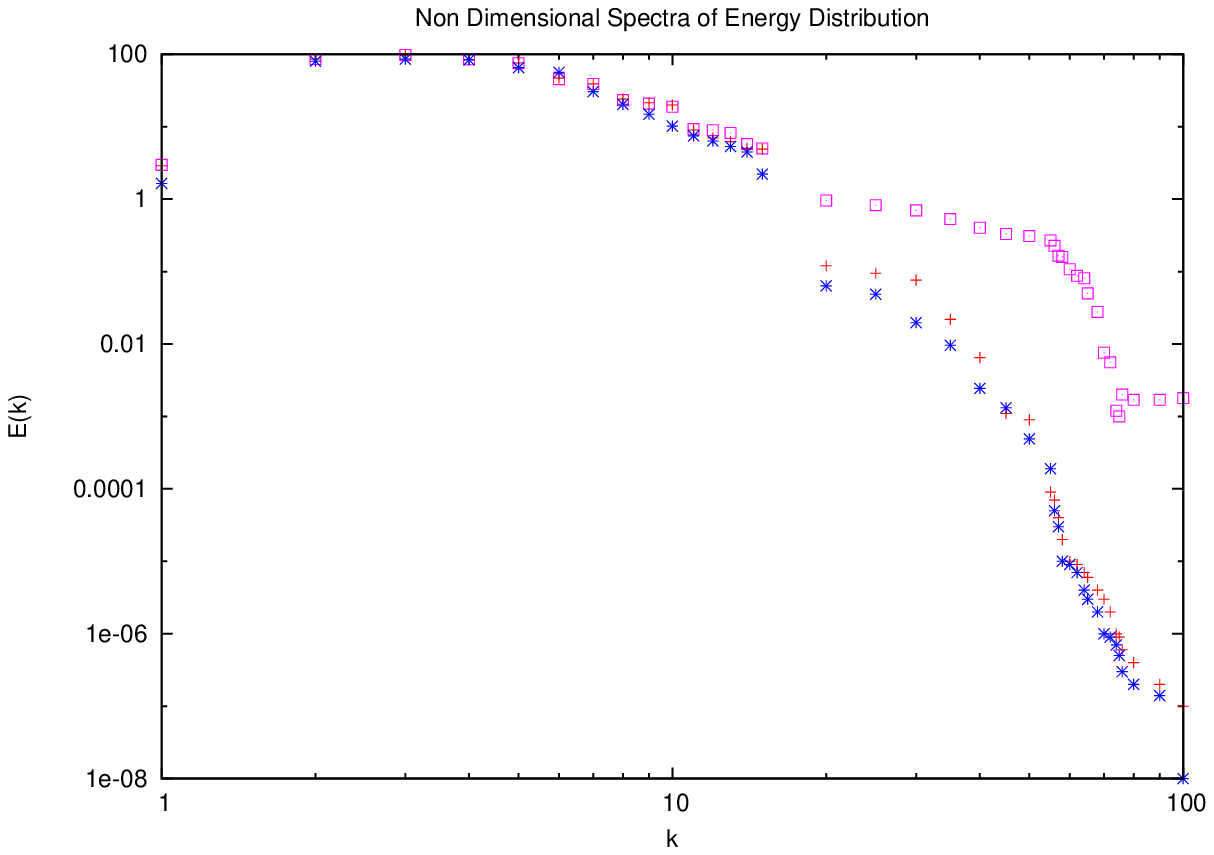}
}
\subfigure[Fig 3. Enstrophy distribution in dispersed phase----with $\times(green)$ and without $\dotplus$  QBP]{ 
\includegraphics[scale=0.45]{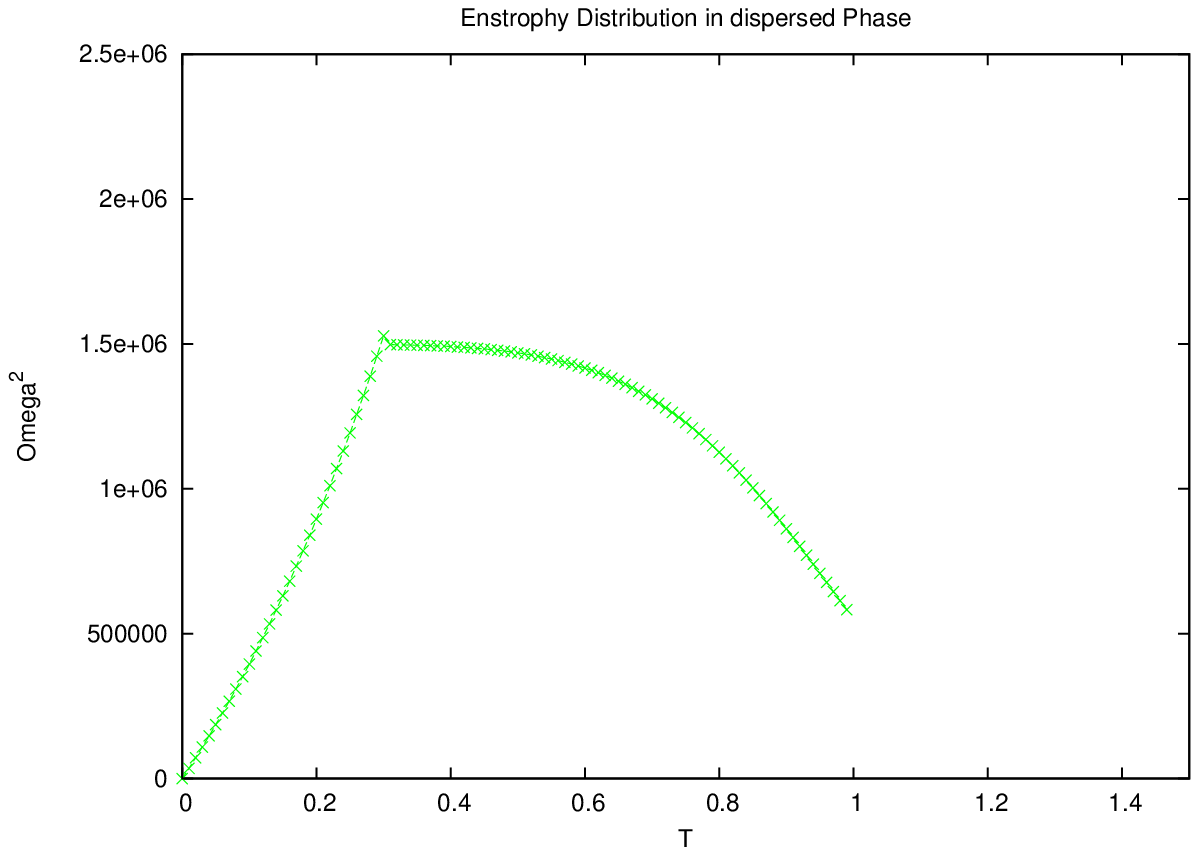}
}
\end{figure}

\section{Particle dynamics and dispersion}
Particle dispersion in Lagrangian simulations is usually measured by tracking individual
particle path and calculating the variance of the relative displacement. In our simulation, 
Eulerian-Eulerian simulation
was performed with one way coupling. We also studied particle dynamics in simulations with gravity 
and without gravity to study the gravitational effect on dynamics with nonlinear force. 
The Eulerian-Eulerian simulation was carried out in a coupled way so that carrier phase turbulence 
affects the dispersed phase through particle dynamics. The interaction in the momentum equation 
is drag force and nonlinear Basset force in the limit of large density ratio $\rho_{\rm p}/\rho_{\rm f} $.\\
The characteristic relaxation time is computed by formulation,
\begin{equation}
 \langle {X_{\rm p}}^{\rm 2} (t) \rangle 
= \cfrac{1}{N} {\sum_{\rm {j=1}}}^{\rm N} {\lbrace
x_{\rm {p,j}}(t) - x_{\rm {p,j}}(t_{\rm 0}) \rbrace}^{\rm 2}
\end{equation}
The characteristic polynomial of the Jacobian Matrix is given by, 
\begin{equation}
(u-\lambda) \left ( (u-\lambda)(u-\lambda) - \cfrac{\gamma P} {\rho} 
\right ) ( \check{u_{\rm p}} -\lambda )
\left ( (\check{u_{\rm p}} -\lambda) (\check{u_{\rm p}} -\lambda) - 
\cfrac{ \cfrac{5}{3} P_{\rm {QB}}}{\check{n_{\rm p}}} 
\right ) = 0
\end{equation}

Particle dispersion can then be related to the time derivative of the quantity \cite{monin}
\begin{equation}
{D_{\rm {p}}}^{\rm L}(t) = \cfrac{1}{2} \cfrac{d}{dt} \langle{X_{\rm p}}^{\rm 2}(t) \rangle
\end{equation} 
Eulerian simulation can not track individual particle path. Particle dispersion is measured 
by semi-empirical method \cite{monin}
If the simulation is being carried out with
colored particles and transport equation is written in terms of ratio of colored 
particles to total particles $(\check{c}=\cfrac{\check{n_{\rm c}}}{\check{n_{\rm p}}}) $. 
Then we can write transport equation as,
\begin{equation}
\cfrac{\partial}{\partial t} \check{c} \check{n_{\rm p}} + \cfrac{\partial}{\partial x_{\rm i} } 
\check{c} \check{n_{\rm p}} \check{u_{\rm {p,i}}} = \cfrac{\partial}{\partial x_{\rm i}} 
\end{equation}
Here $ {\check{u_{\rm {p,i}}}}^{\rm c}$ is the mesoscopic velocity of colored particles. 
Since only the velocity of total particle is resolved, the right hand side term takes into account 
of the slip velocity between colored and mesoscopic velocity of the particle ensemble. 
Comparing to Navier-Stokes equation, this term is equivalent to molecular diffusion. Since slip velocity arises only 
from uncorrelated movement of particles, this term can be modeled as diffusion term related to 
quasi brownian motion.
If the ensemble averaged mean number density fraction of colored particles 
$ \langle \check{n_{\rm p}}\rangle C=  \langle \check{n_{\rm p}}\check{c} \rangle $ 
is uniformly stratified in the $ \it k$ direction[here in $-k$ direction] $ \check{c} =
C + \acute{c} $ and  fluctuations are assumed periodic with respect to the computer
domain, then fluctuation number density of colored particles
 $ \acute{c} \tilde{n_{\rm p}} $ can be calculated  from the total colored particle 
density function. One obtains a transport equation for the fluctuation of colored particle 
concentration as, 
\begin{equation}
\cfrac{\partial}{\partial t} \acute{c} \check{n_{\rm p}} + 
\cfrac{\partial}{\partial x_{\rm i}} \acute{c} \check{n_{\rm p}} \check{u_{\rm {p,i}}} =
-\check{n_{\rm p}} \check{u_{\rm {p,k}}} \cfrac{\partial}{\partial x_{\rm k}}C 
+ \cfrac{\partial}{\partial x_{\rm i}} \check{c} \check{n_{\rm p}} (\check{u_{\rm {p,i}}} -
{\check{u_{\rm {p,i}}}}^{\rm c} )
\end{equation}
Averaging the colored number density equation (15), one obtains Reynold
averaged type transport equation, 
\begin{equation}
\cfrac{\partial}{\partial t} \langle \check{n_{\rm p}} \rangle C 
+ \cfrac{\partial}{\partial x_{\rm i}} \langle \check{n_{\rm p}} \rangle C
\langle \check{u_{\rm {p,i}}} \rangle_{\rm p} = - \cfrac{\partial}{\partial x_{\rm i}}  
\langle \check{n_{\rm p}} \acute{c} u_{\rm {p,i}} \rangle + \cfrac{\partial}{\partial x_{\rm i}} 
\langle \check{c} \check{n_{\rm p}} ( \check{u_{\rm {p,i}}} -{\check{u_{\rm {p,i}}}}^{\rm c} ) \rangle
\end{equation}
Particle dispersion term can be derived by making gradient assumption,
\begin{equation}
(\langle \acute{c} \check{n_{\rm p}} \check{u_{\rm {p,k}}} \rangle = \langle \check{n_{\rm p}}
{D_{\rm {p,k}}}^{\rm t} \cfrac{\partial}{\partial x_{\rm k}} C \rangle 
\end{equation}
A semi empirical diffusion coefficient can be defined as, 
\begin{equation}
{D_{\rm {p,k}}}^{\rm t} = \cfrac{ \langle 
\check{n_{\rm p}}  \acute{c} u_{\rm {p,k}} \rangle } {\langle \check{n_{\rm p}} \rangle 
\cfrac{\partial}{\partial x_{\rm k}} C } 
\end{equation}
This dispersion coefficient is comparable to the Lagrangian Dispersion coefficient(13) 
in the long time limit of stationary turbulence. Simulations without QBP likely underestimates 
Lagrangian dispersion. 
The characteristic particle relaxation time is computed by the formulation given by, 
\begin{equation}
\tau_{\rm p} = \cfrac{ \rho_{\rm p} d^{\rm 2} } { 18f(Re_{\rm p} ) {\mu} } 
\end{equation}
Particle Reynold number for the drag force correction $f(Re_{\rm p})$ is based  
\begin{equation}
f(Re_{\rm p}) = 1 + 0.15{Re_{\rm p}}^{\rm {0.687d0}} 
\end{equation}

\section{Numerical Simulations and results}
Particle dynamics and particle dispersion have been studied by experiments and Lagrangian computations.
Experimental results of Snyder \& Lumley \cite{snyder}(referred as SL) is very robust test case for numerical 
simulation. They inserted particles with different inertial properties
into grid generated spatially decreasing turbulence and measured particle
dynamics as well as particle dispersion. The test case has been computed with Lagrangian approach 
by Elghobashi \& Truesdell(ET) \cite{elghobashi} . The carrier phase was taken as 
a temporarily decreasing homogeneous isotropic turbulence corresponding to the grid 
generated turbulence of SL. After an initial calculations for two turnover time
$ (t=\cfrac{l_{\rm {ii}}} {\acute{u_{\rm f}}} )$,  particles were inserted into the flow. Particle dynamics and
dispersion were analyzed by ET \cite{elghobashi} on particles similar to the experiments of 
Snyder \& Lumley \cite {snyder}
for direct comparison with lagrangian simulation.
We carried out particle simulation with Eulerian-Eulerian
approach and comparison with the experimental results and Lagrangian simulation results were attempted. 
The simulations were performed on $100^{\rm 3}$ grid. \\
The carrier phase velocity is initialized at dimensionless time T=0 with 
a divergence velocity field(continuum condition) such that the kinetic energy 
satisfies the spectrum \cite{elghobashi}
\begin{equation}
E(k,0) = \cfrac{3}{2}{\acute{u_{\rm {f,0}}}^{\rm 2}} \cfrac{k}{k_{\rm p}} \exp((-k/k_{\rm p}) 
\end{equation}
where $\acute{u_{\rm f}} $ is the dimensionless rms velocity, k is the wave number 
and $k_{\rm p}$ is wavenuber of peak energy. All wave number were normalized with $k_{\rm {min}}$.
In the present simulation, properties of carrier phase was validated against the 
properties of carrier phase turbulence of SL and ET. The spatial evolution of the flow
in the experiment of SL is converted to a temporal evolution of the flow
by $t=\cfrac{x}{\check{U}}$ . here $\check{U}$ is the mean 
convection velocity in the experiment. \\ 
Figure 4 shows  dimensionless velocity square  in the carrier phase in comparison 
with Lagrangian simulation(ET) and experiment(SL)
results. 
\begin{figure}[ht]
\subfigure[Fig 4. Squared dimensionless velocity Distribution--Lagrangian ($\times$)  
and present(---)  simulation]{
\includegraphics[scale=0.50]{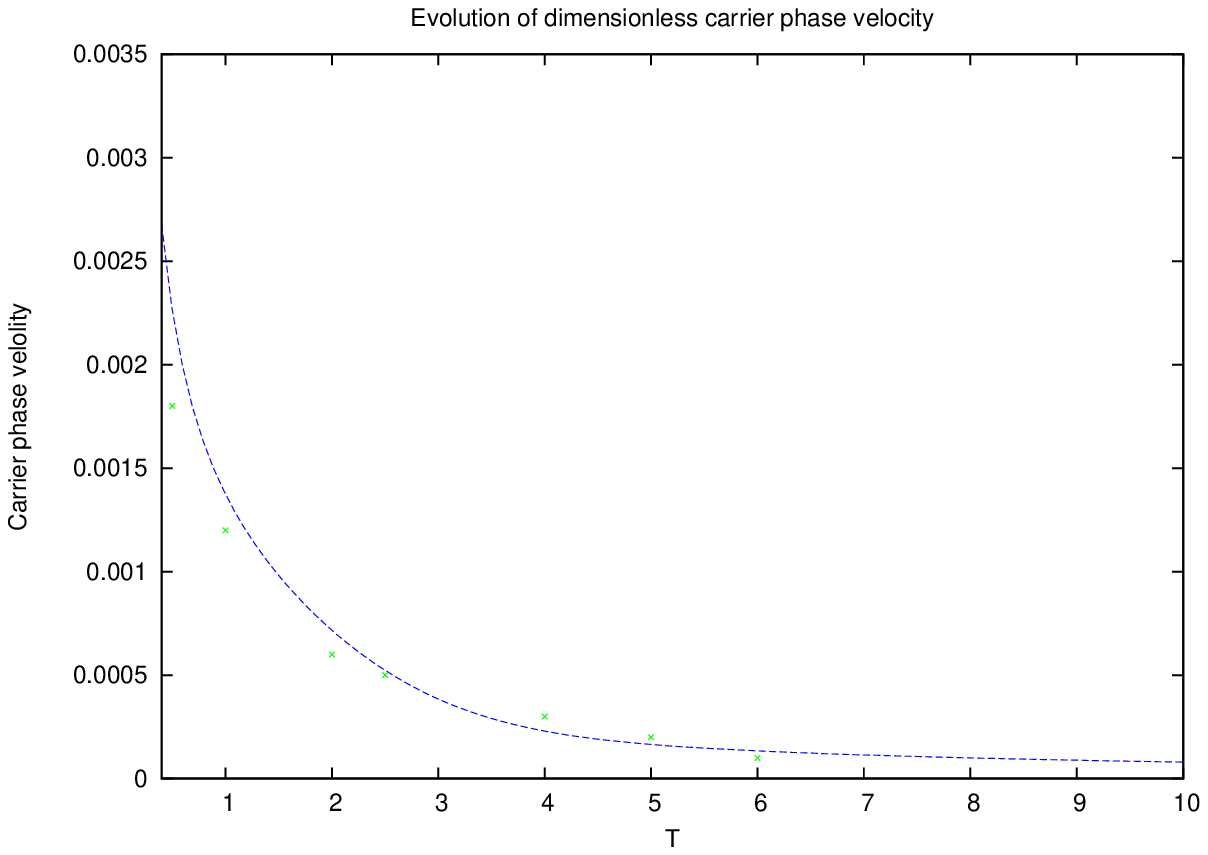}
}
\subfigure[Fig 5. Comparison of dissipation energy in carrier phase with Lagrangian simulation]{
\includegraphics[scale=0.50]{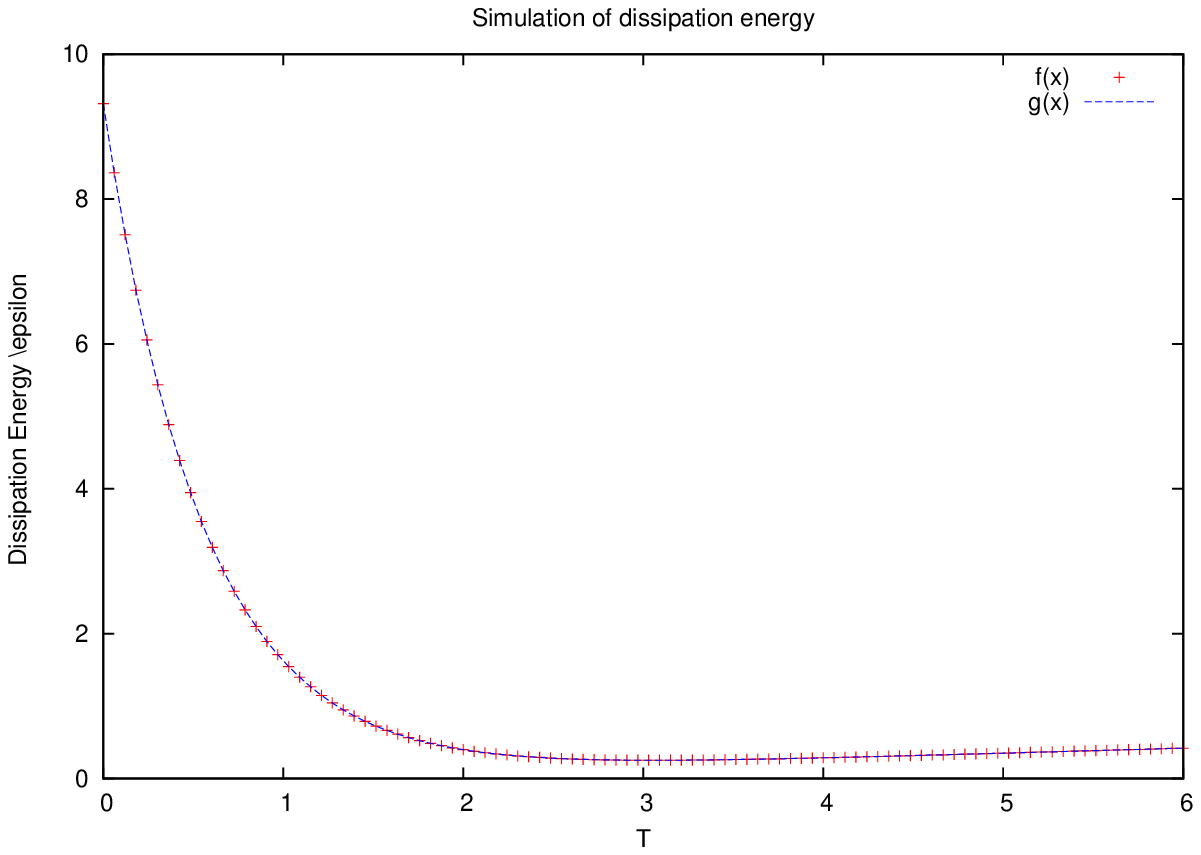} 
}
\end{figure}
To verify numerical simulation resolution, dissipation energy $\epsilon$ is also compared to the 
temporal change of 
kinetic energy $\cfrac{d}{dt} {q_{\rm f}}^{\rm 2}$ in Figure 5. Our result shows excellent agreement 
between dissipation 
and kinetic energy decrease. So it can be assumed that 
numerical dissipation is negligible compared to viscous dissipation. \\
In Figure 6, Reynold number from our simulation with Lagrangian simulation. The present 
simulation more rapid decay of turbulent Reynold number compared to the Lagrangian simulation(ET). 
\begin{figure}[ht]
\subfigure[Fig 6. Evolution of Reynold Number in carrier phase $\times(green)$ Lagrangian and $\dotplus(red)$ present simulation]{
\includegraphics[scale=0.50]{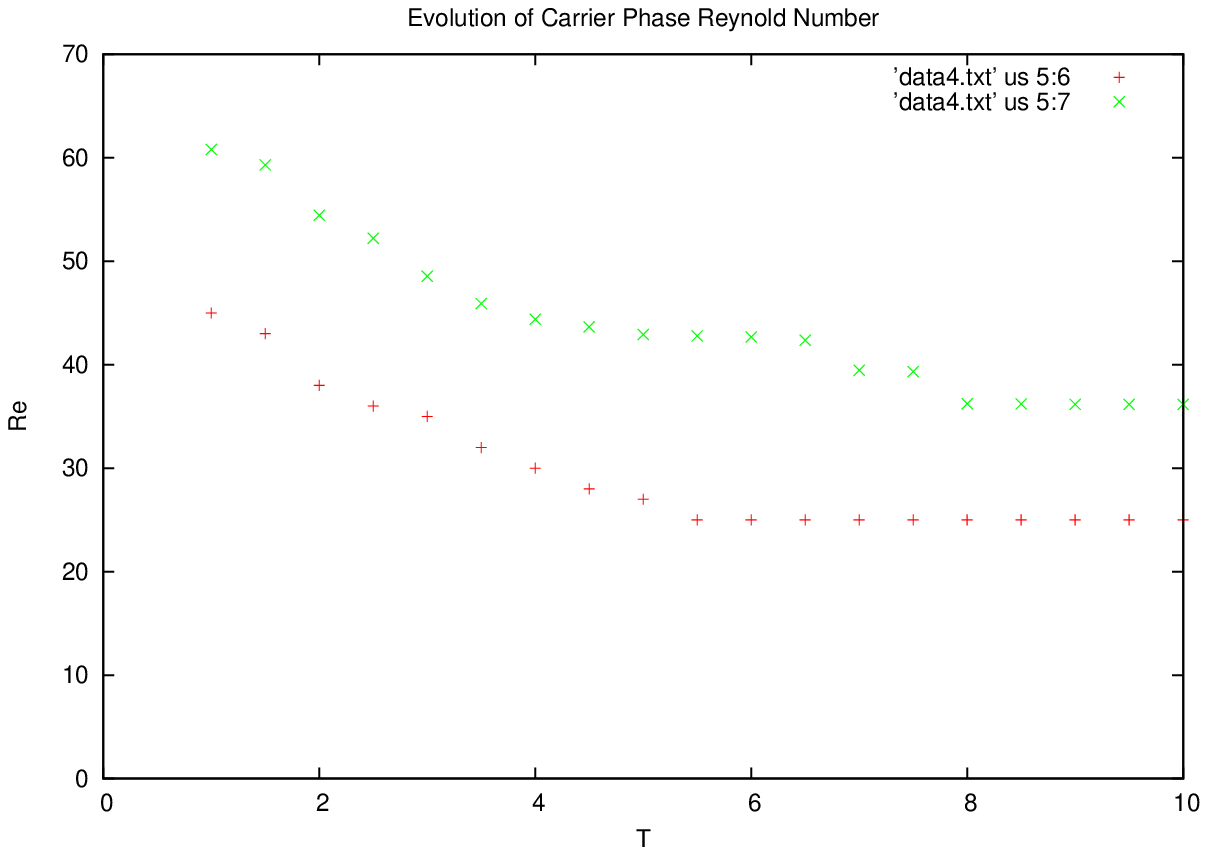}
}
\subfigure[Fig 7. Evolution of integral length scale $\times(pink)$..Lagrangian and $\dotplus(green)$ ..present simulation]{
\includegraphics[scale=0.50]{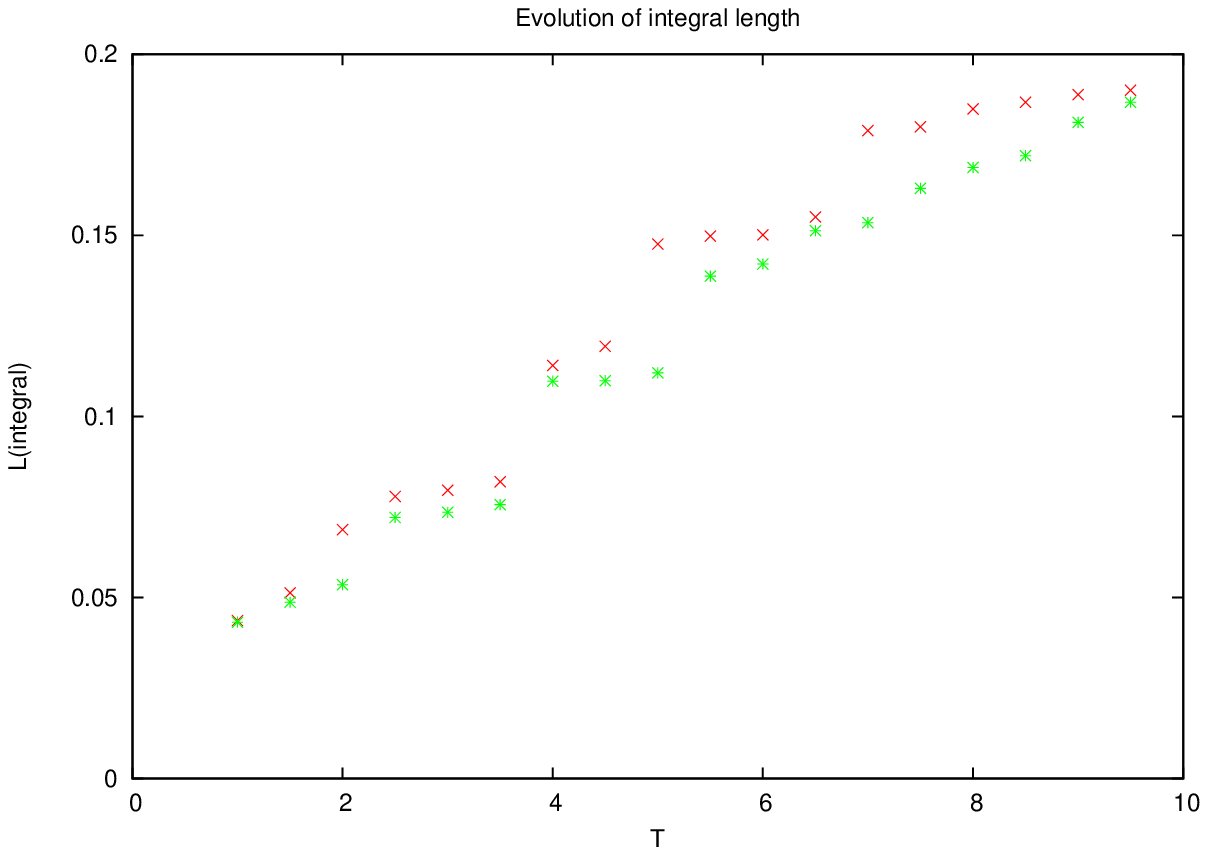}
}
\end{figure}
This is because the integral length scale increase slowly in our simulation(Figure 7). 
The temporal evolution of integral time length scale shows similar qualitative behavior in eulerian simulation.
Particles were inserted as in the Lagrangian simulation $ {{u_{\rm f}}^{\rm \prime}}^{\rm 2}$ 
(ET) at nondimensional time $ T=2.0$ with same velocity as carrier phase when inserted into the turbulent flow. 
Particle properties were analyzed in turbulence with and without gravity. When particles are inserted into
gravity, they establish a mean terminal velocity vertically downward direction. The gravity constant was
calculated as in the experiment. For all types of particle sizes, relative square velocity in the present 
simulation shows same qualitative behavior as in the Lagrangian simulation. 
Figure 8 and 9  shows the squared relative  velocity distribution of the carrier phase without and 
with the presence of gravity. The gravity constant is measured such that same ratio of 
$ \cfrac{v_{\rm {t,0}}}{u_{\rm 0}}$. 
In all cases, present simulation shows same behavior as Lagrangian simulation. 
We have chosen temporarily
decaying turbulence condition in our simulation as initial parameter.   
In Eulerian simulations, one does not have access to individual particle paths.
Particle dispersion can still be measured by a semi-empirical
method \cite{monin}. 
Since only the velocity of the total droplet number is resolved which brings another 
extra term on right hand side of the above equation. 
This term takes into account the slip velocity between colored particle and the 
mesoscopic velocity of particle ensemble. \\
 
dispersed phase in excellent agreement. 
It is observed in the present simulation that when gravity is taken into account, 
particle dynamics was modified. The observed crossing trajectory effect is due to mean settling
velocity of the particle and leads to decrease in integral time scale of fluid turbulence. 
Such an effect leads to an increase of effective particle stokes number. This also increases 
the relative squared velocity value with respect to the non settling value. 
 Particle dispersion is also calculated for the dispersed phase. 
In order to compare with the carrier phase, equation (16) is solved for carrier phase 
without molecular diffusion.  In the Eulerian simulation with gravity taking into account, particle 
dispersion is significantly lower than the simulation without gravity consistent with 
Csanady analysis \cite{csanady}. Calculation for all particles show that Eulerian 
simulation shows similar qualitative behavior as Lagrangian simulation and dispersion calculation 
is quantitatively lower than Lagrangian simulation. 
\begin{figure}[ht]
\subfigure[Fig 8. Evolution relative velocity(dimensionless) with gravity] {
\includegraphics[scale=0.50]{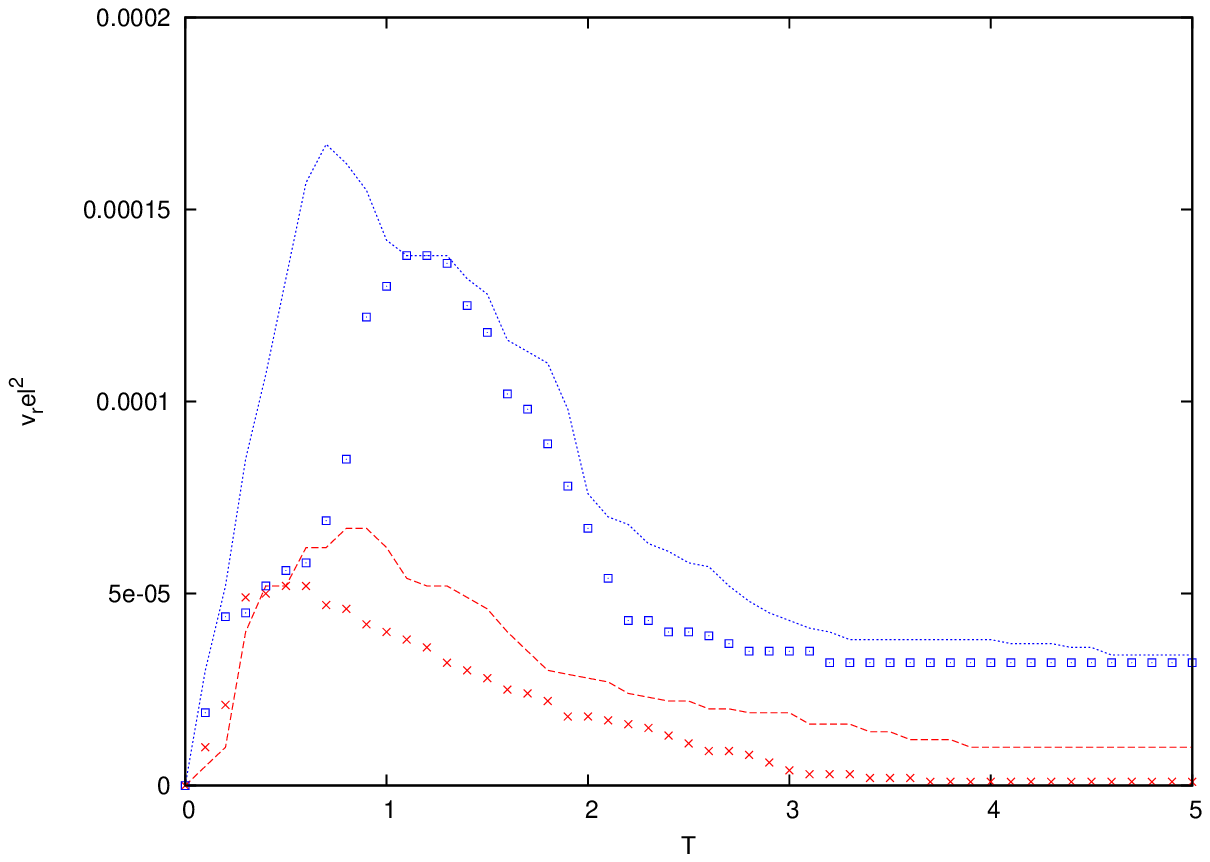}
}
\subfigure[Fig 9. Evolution of relative velocity(dimensionless) without gravity]{
\includegraphics[scale=0.50]{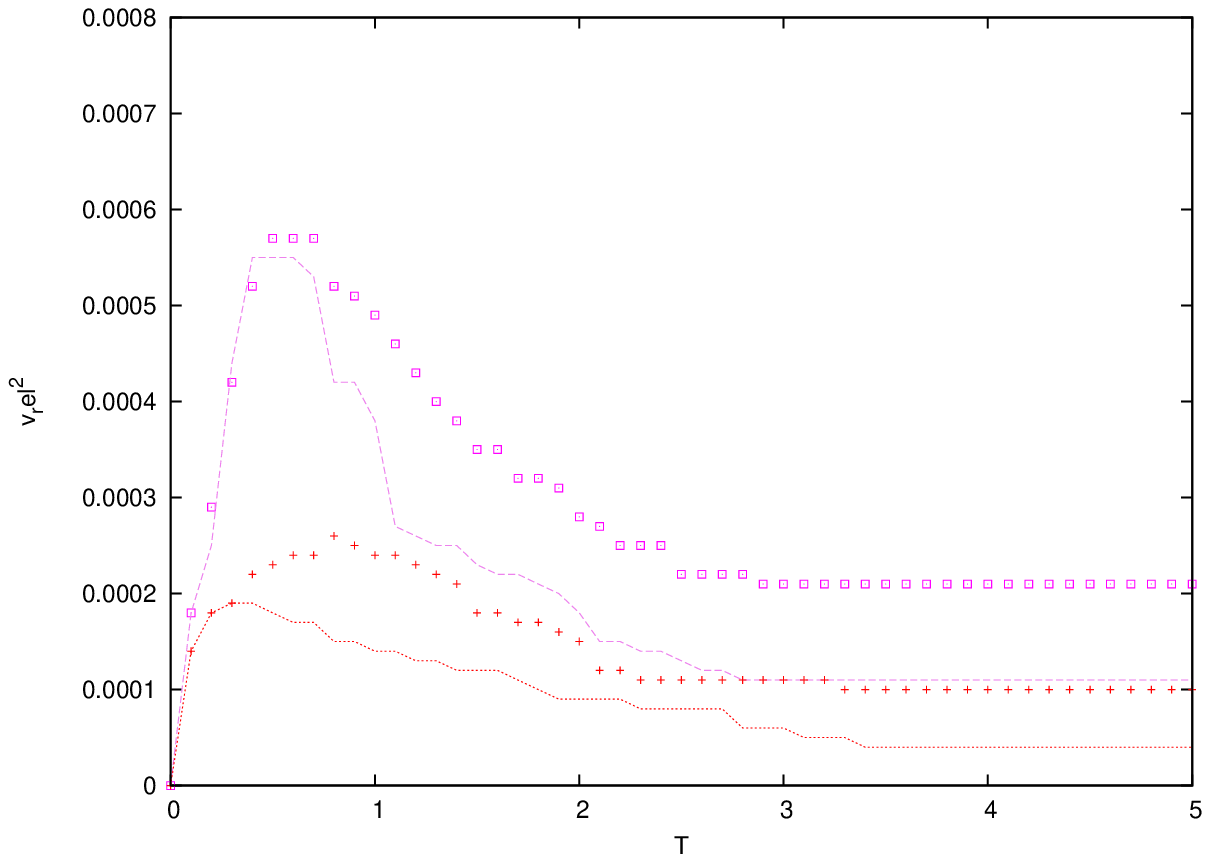}
}
\end{figure}

The Eulerian equations for the dispersed phase have been implemented on the slip velocity
\begin{equation}
Re_{\rm p} = \cfrac{ \vert { \vec{u_{\rm p}} - \vec{u_{\rm f}}  \vert }d }{\nu_{\rm f}}
\end{equation}

\begin{figure}[ht]
\subfigure[Fig 10. Evolution of dispersion coefficient without gravity]{
\includegraphics[scale=0.50]{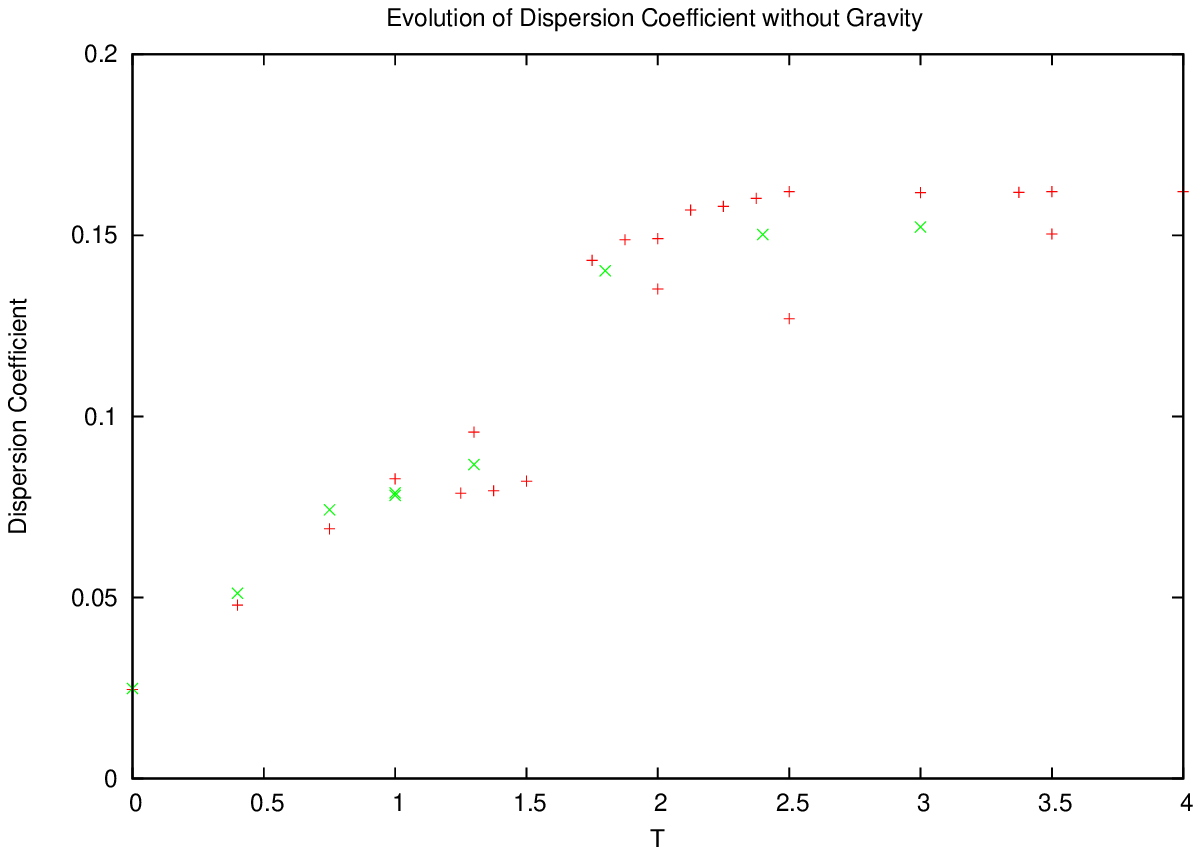}
}
\subfigure[Fig 11. Evolution of dispersion coefficient with gravity]{
\includegraphics[scale=0.50]{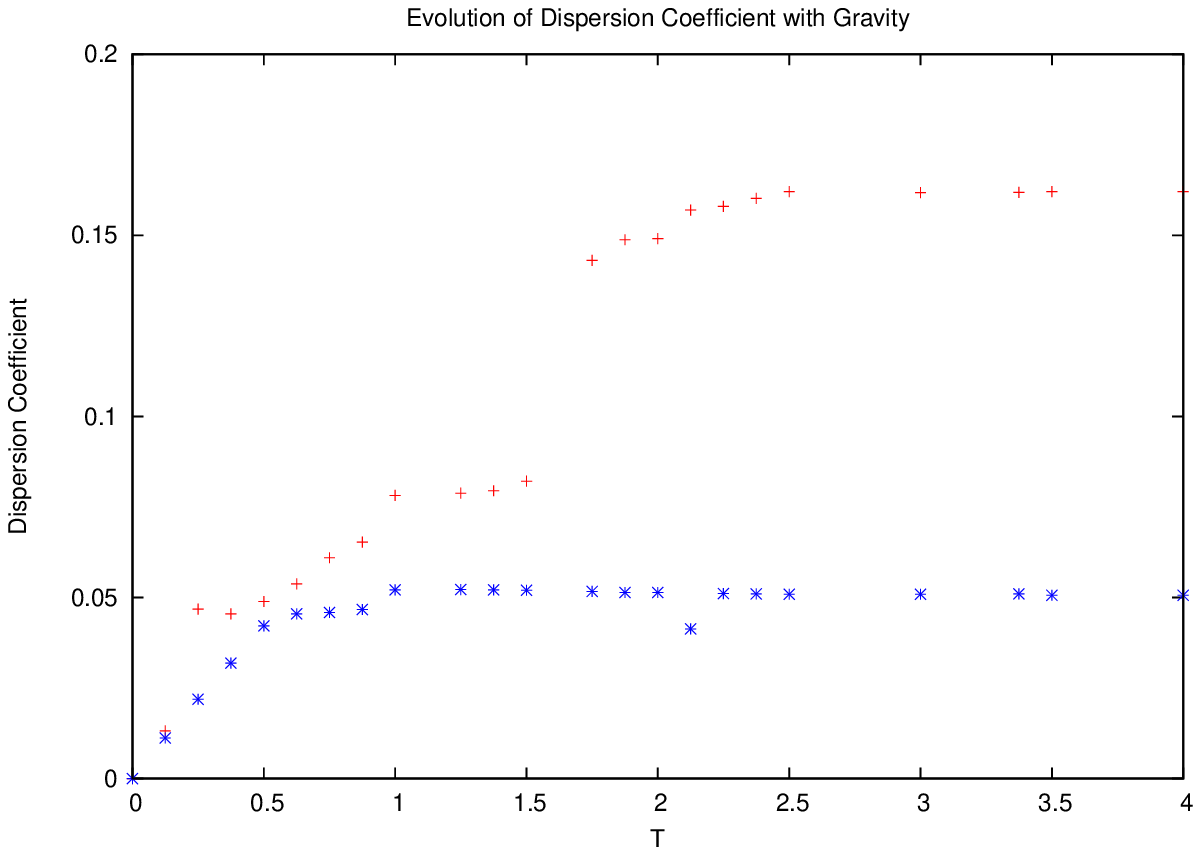}
}
\end{figure}
The Eulerian mean square relative velocity differs from the Lagrangian mean square 
velocity  by the quantity $ \delta {u_{\rm p}}^{\rm 2} $ from QBM.
Figure 12 shows the temporal development of the carrier phase $ {u_{\rm f}}^{\rm 2} $. 
Balakin et al \cite{balakin} also studied uniformly sized sediment particle movements in viscous fluids using 
Eulerian-Eulerian simulation and compared with experimental results 
of Nicolai et. al \cite{nicolai} for sedimenting particles.  
They corresponding governing equation is given by,  
\begin{equation}
\cfrac{\partial (\alpha_{\rm m} \rho_{\rm m}) }{ \partial t} + 
\bigtriangledown(\alpha_{\rm m}\rho_{\rm m} \vec{u_{\rm m}})=0 
\end{equation}
Index ${\it m} $ assigns the phase(liquid or solid), $\alpha$ is the volume fraction, $\rho$ 
is the density and $\vec{u}$ is mean phase velocity. The momentum equation is given by, 
\begin{equation}
\cfrac{\partial (\alpha_{\rm m} \rho_{\rm m} \vec{u_{\rm m}})}{\partial t} 
+ \bigtriangledown(\alpha_{\rm m} \rho_{\rm m} \vec{u_{\rm m}} \vec{u_{\rm m}} ) = - \alpha_{\rm m}
\vec{\bigtriangledown} p + \alpha_{\rm m} \rho_{\rm m} \vec{g} 
 + \bigtriangledown(\alpha_{\rm m} \tau_{\rm m}) + \vec{M_{\rm m}} + \vec{F_{\rm A}}
\end{equation}
In further to investigate our simulation approach, we carried out simulation for 
sedimenting particle flow simulation similar to the experiment done by Nicolai et al. \cite{nicolai} 
Figure 11 shows the comparison between our simulation with Experimental results of Nicolai etl al. \cite{nicolai} and 
also Balakin et al \cite{balakin}
simulation results. 
The dispersed phase velocity is plotted against volume fraction of carrier phase to dispersed phase. 
Our simulation strongly agree with experimental data.
\begin{figure}[ht]
\subfigure[Fig 12: Time average settling velocity plotted as a function of solid volume fraction, red $\times \cdots $ (Nicolai et al), 
blue  $\times \cdots$ balakin et al and green $\circ \cdots $ (our simulation)]{
\includegraphics[scale=0.50]{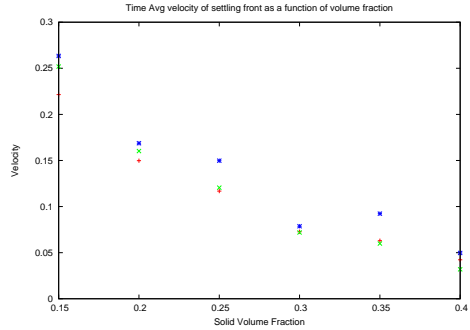}
}
\end{figure}.   

\section{Conclusion}
Particle dispersion is measured for the dispersed phase. In order to compare with the carrier phase, 
the equivalent of equation(16) is solved for carrier phase without molecular diffusion. 
A preliminary model of Quasi Brownian Motion was used to study unresolved kinetic particle energy.
This model allowed simulations to compare with the experimental results of Snyder \& Lumley \cite{snyder}. We also have done for 
sedimenting particles similar to experiment done by Nicolai et. al \cite{nicolai} and compared with Balakin simulation results.
\cite{balakin} which strongly shows that Quasi Brownian ensemble approach is very strong alternative tool for two phase flow modeling.

\end{document}